\documentclass[prl,twocolumn,showpacs,superscriptaddress,floatfix]{revtex4}
\usepackage{graphicx,amssymb,amsmath}

\newlength{\ldag}
\newcommand{\cdag}{c^\dagger}
\newcommand{\anc}{c^{\phantom\dagger\hspace{-\ldag}}}
\newcommand{\adag}{a^\dagger}

\newcommand{\bdag}{b^\dagger}

\newcommand{\rmR}{\mathrm{R}}

\newcommand{\ve}{\varepsilon}

\begin{document}

\title{Finite-size scaling exponents and entanglement in the two-level BCS model}

\author{S\'ebastien Dusuel}
\email{sdusuel@thp.uni-koeln.de}
\affiliation{Institut f\"ur Theoretische Physik, Universit\"at zu
K\"oln, Z\"ulpicher Str. 77, 50937 K\"oln, Germany}

\author{Julien Vidal}
\email{vidal@lptmc.jussieu.fr}
\affiliation{Laboratoire de Physique Th\'eorique de la Mati\`ere Condens\'ee, CNRS UMR 7600,
Universit\'e Pierre et Marie Curie, 4 Place Jussieu, 75252 Paris Cedex 05, France}

\begin{abstract}

We analyze the finite-size properties of the two-level BCS model. Using the continuous unitary transformation technique, we show that nontrivial scaling exponents arise at the quantum critical point for various observables such as the magnetization or the spin-spin correlation functions. We also discuss the entanglement properties of the ground state through the concurrence which appears to be singular at the transition.

\end{abstract}

\maketitle



Since its experimental discovery in 1911 by Kamerlingh Onnes, superconductivity has been the object of intensive research. More than 45 years elapsed before Bardeen, Cooper and Schrieffer (BCS) gave it a theoretical foundation \cite{Bardeen57}. The revival of interest for superconductivity in the last two decades originates mainly in the inability of the BCS theory to explain neither high-$T_\mathrm{c}$ superconductivity nor finite-size effects in nanoscale grains \cite{vonDelft01}. The effect of discreteness in the energy spectrum of nanograins has been studied in the reduced BCS model [see Eq.~(\ref{eq:ham_bcs_fermions})], whose exact solution was obtained by Richardson in 1963 \cite{Richardson63_1,Richardson63_2,Richardson64} and whose integrability was proved only recently \cite{Cambiaggio97}. 

In this Communication, we focus on the two-level reduced BCS model, which displays a second-order quantum phase transition to a superconducting state, for a finite value of the electronic attraction. 
At the critical point, we show that the spectrum and correlation functions possess nontrivial finite-size scaling exponents as already suggested for the ground-state energy \cite{Roman02,Roman04}. 
Following the same line as for the Lipkin-Meshkov-Glick (LMG) model  \cite{Dusuel04_3,Dusuel04_4}, we combine a $1/N$ expansion,  the continuous unitary transformations (CUTs) technique,  and a scaling argument, to exactly determine these exponents. Our results are supported by a numerical investigation of the finite-size effects.
In a second step, we discuss the entanglement properties of the ground state via the so-called concurrence \cite{Wootters98}. Using a standard mapping of the reduced BCS model onto a  spin system, we show that,  as in one-dimensional spin chains  \cite{Osborne02,Osterloh02}, this concurrence displays some singular behavior at the transition point. In the present case, one may, however, distinguish between two cases depending on which level the two spins considered for the concurrence belong to.


We consider the reduced BCS Hamiltonian \cite{Bardeen57,vonDelft01}
%
\begin{equation}
  \label{eq:ham_bcs_fermions}
  H=\frac{1}{2}\sum_{i,\sigma=\pm} \ve_i \cdag_{i,\sigma}\anc_{i,\sigma}
  -\frac{g}{N}\sum_{i,j} \cdag_{i,+} \cdag_{i,-} \anc_{j,-} \anc_{j,+},
\end{equation}
%
where $\cdag_{i,\pm}$ and $\anc_{i,\pm}$ are fermionic creation and
annihilation operators obeying the anticommutation relation
$[\anc_{i,\sigma},\cdag_{i',\sigma'}]_+=\delta_{i,i'}\delta_{\sigma,\sigma'}$,
$g$ is a positive coupling constant that we set to unity,
$N$ is the total number of states ($j=1,\ldots, N$) that is assumed to be
even, and the $1/N$ factor ensures that the thermodynamical limit is well behaved.
For the sake of simplicity, we furthermore restrict ourselves to the subspace
where all states $j$ have occupation number 0 or 2. This subspace is not
coupled to its complement by the BCS Hamiltonian.
Via the mapping 
%
\begin{equation}
  \label{eq:mapping_fermions_spins}
  \sigma_+^i=\anc_{i,-} \anc_{i,+}, \quad
  \sigma_z^i=1-\cdag_{i,+} \anc_{i,+}-\cdag_{i,-} \anc_{i,-},j
\end{equation}
%
of fermionic to spin operators, the two-level Hamiltonian
(\ref{eq:ham_bcs_fermions}) at half-filling becomes
%
\begin{equation}
  \label{eq:ham_bcs_spins}
  H=-h(S_{1z}-S_{2z})-\frac{1}{N}\left( S_x^2+S_y^2\right).
\end{equation}
%
We have denoted $\pm h$ the energies of the two levels ($h\geq 0$),
which are interpreted as magnetic fields in the spin language.
We have introduced total spin operators
$S_{1(2)\alpha}=\sum_{i\in I_{1(2)}} \sigma_\alpha^i/2$,
$I_{1(2)}$ being the set of states $i$ with energy $+h$ ($-h$),
and $S_\alpha=S_{1\alpha}+S_{2\alpha}$ (where $\alpha=x,y,z$).

Note that we have also used the conservation of the number of fermions,
which reads $S_z=0$ in the spin language.
The Hamiltonian (\ref{eq:ham_bcs_spins}) further commutes with the
total spin operators ${\bf S}_1^2$ and ${\bf S}_2^2$. Here, we focus on the maximum spin sector to which the 
low-energy states belong to.


In the spin language, the two-level BCS Hamiltonian thus describes the
physics of two coupled $XX$ models with infinite range (constant) 
interactions, and embedded in transverse magnetic fields of 
opposite directions. 
In the thermodynamical limit, the Hamiltonian (\ref{eq:ham_bcs_spins})
undergoes the well-known BCS mean-field quantum phase transition at $h=1$,
that we now briefly describe. 
Replacing the large spin operators by their classical value ($k=1,2$)
%
\begin{equation}
  \label{eq:classical_spin}
  {\bf S}_k=\frac{N}{4}\big(\sin \theta_k \cos \phi_k , \sin \theta_k\sin\phi_k, \cos \theta_k \big),
\end{equation}
%
and minimizing (\ref{eq:ham_bcs_spins}) with respect to  $\theta$ and $\phi$, yields the solution
$(\theta_1^*=0,\theta_2^*=\pi)$ in the symmetric phase $h\geq 1$, and
$(\theta_1^*=\arccos h,\theta_2^*=\pi-\theta_1^*)$ in the broken phase $h<1$.
The spontaneous breaking of the rotational symmetry along $z$ for $h<1$
gives rise to infinitely degenerate ground states
($\phi_1^*=\phi_2^*$ can take any value).


%
\begin{table}[t]
  \centering
  \begin{tabular}{|c|c|c|c|}
    \hline
    $\Phi$ & $\xi_\Phi$ & $n_\Phi$ & $n_\Phi+2\xi_\Phi/3$\\
    \hline
    \hline
    $e_0$ & 1/2 & 1 & 4/3\\
    \hline
    $\Delta$ & 1/2 & 0 & 1/3\\
    \hline
    $4\langle S_{1 z} \rangle/N$ & -1/2 & 1 & 2/3\\
    \hline
    $16\langle S_{1z}^2 \rangle/N^2$ & -1/2 & 1 & 2/3\\
    \hline
    $16\langle S_{1x} S_{2x}\rangle/N^2$ & -1/2 & 1 & 2/3\\
     \hline
  \end{tabular}
  \caption{Scaling exponents for the ground-state energy per spin $e_0$, the gap $\Delta$, the magnetization 
  $\langle S_{1 z} \rangle$, and the two-point correlation function $\langle S_{1z}^2 \rangle$ and 
  $\langle S_{1x} S_{2x}\rangle$.}
  \label{tab:exponents}
\end{table}

We compute the finite-size scaling exponents following the same strategy as 
in Refs.~\onlinecite{Dusuel04_3} and \onlinecite{Dusuel04_4}. 
We consider the symmetric phase $h\geq 1$, and use the bosonic 
Holstein-Primakoff representation \cite{Holstein40} of both spins around the 
mean-field ground state
%
\begin{eqnarray}
  \label{eq:HP1}
  S_{1z} &=& N/4-\adag a, \quad S_{2z} = \bdag b-N/4,\\
  \label{eq:HP2}
  S_{1+} &=& (S_{1-})^\dagger=(N/2)^{1/2}\left(1-2\adag a/N\right)^{1/2}a,\\
  \label{eq:HP3}
  S_{2+} &=& (S_{2-})^\dagger=(N/2)^{1/2}\bdag\left(1-2\bdag b/N\right)^{1/2}.
\end{eqnarray}
%
One then inserts these in Eq.~(\ref{eq:ham_bcs_spins}), and expands the square 
roots to the lowest order needed to compute the exponents, i.e., $(1/N)^1$. 
The Hamiltonian is not diagonal and contains terms creating or destroying 
one $a$ and one $b$ excitations, for example $\adag\bdag+ab$. 
Since it is quartic,  it  cannot be simply diagonalized by a 
Bogoliubov transform, but one can use the CUTs 
\cite{Wegner94,Glazek93,Glazek94} to perform this task 
(see also Refs.~\onlinecite{Dusuel04_2} and \onlinecite{Dusuel04_4} 
for details). 
Starting from the initial Hamiltonian $H(0)=H$, 
one considers a unitary equivalent Hamiltonian $H(l)$ satisfying 
the flow equation
%
\begin{equation}
  \label{eq:dlH}
  \partial_l H(l)=[\eta(l),H(l)].
\end{equation}
%
$\eta(l)$ is the generator of the unitary transformation, chosen to make the 
final Hamiltonian $H(l=\infty)$ diagonal. For the problem at hand, 
the simplest generator is the so-called particle-conserving generator 
\cite{Knetter00}, with particle number operator $Q=n_a+n_b$, 
where $n_a=\adag a$ and $n_b=\bdag b$. With this choice, 
the final Hamiltonian is polynomial in $n_a$ and $n_b$. 
To compute correlation functions, one has to follow the flow of spin 
operators. $S_z=n_b-n_a$ is found to have no flow, 
so that eigenstates of the final Hamiltonian must satisfy $n_a=n_b$ 
to fulfill $S_z=0$. The ground state (first excited state) of $H(l=\infty)$ 
is the state with $n_a=0(1)$ boson. 

Concerning the spectrum, we focus on the ground-state energy per site 
$e_0$ and the gap $\Delta$. For the magnetization, one has
$\langle S_{2 z} \rangle=-\langle S_{1 z} \rangle$, and the rotation 
invariance around the $z$ axis implies 
$\langle S_{k x} \rangle=\langle S_{k y} \rangle=0$ for $k=1,2$. 
Finally, all spin-spin correlation functions 
$\langle S_{j \alpha} S_{k \beta}+S_{k \beta} S_{j \alpha}\rangle$ with 
$j,k \in \{1,2\}$ and $\alpha, \beta \in \{x,y,z\}$ either vanish or can 
be deduced from $\langle S_{1z}^2 \rangle$ and $\langle S_{1x} S_{2x}\rangle$.
As in our earlier works on the LMG model \cite{Dusuel04_3,Dusuel04_4}, 
the flow equations can be integrated exactly, and the five quantities of 
interest are found to behave as 
%
\begin{eqnarray}
  \label{eq:expansion_e0}
  &&e_0 = -\frac{h}{2}+\frac{1}{N}\left(-h+\Xi(h)^{1/2}\right)\\
  &&\hspace{2.5cm}+\frac{1}{N^2}\left[-\frac{h(2h-1)}{2\Xi(h)}
    -\frac{h}{\Xi(h)^{1/2}}\right],\nonumber\\
  \label{eq:expansion_Delta}
  &&\Delta = 2\Xi(h)^{1/2}+\frac{1}{N}\left[ \frac{h(4h-1)}{\Xi(h)}
    -2\frac{h}{\Xi(h)^{1/2}}\right],\\
  \label{eq:expansion_S1z}
  &&\frac{4\langle S_{1 z}\rangle}{N} = 1
  +\frac{1}{N}\left[2-\frac{2h-1}{\Xi(h)^{1/2}}\right]\\
  &&\hspace{2.5cm}+\frac{1}{N^2}\left[ \frac{h^2}{\Xi(h)^2}
    -\frac{h}{\Xi(h)^{3/2}}\right],\nonumber\\
  \label{eq:expansion_S1z2}
  &&\frac{16\langle S_{1z}^2 \rangle}{N^2} = 1
  +\frac{1}{N}\left[4-\frac{2(2h-1)}{\Xi(h)^{1/2}}\right]\\
  &&+\frac{1}{N^2}\left[ \frac{2h(4h^3-8h^2+6h-1)}{\Xi(h)^2}
    -\frac{2h(4h^2-6h+3)}{\Xi(h)^{3/2}}\right],\nonumber\\
  \label{eq:expansion_S1xS2x}
  &&\frac{16\langle S_{1x} S_{2x}\rangle}{N^2} = 
  \frac{1}{N}\frac{1}{\Xi(h)^{1/2}}\\
  &&\hspace{2.5cm}+\frac{1}{N^2}\left[ -\frac{h(2h-1)^2}{\Xi(h)^2}
    +\frac{h(4h-3)}{\Xi(h)^{3/2}}\right],\nonumber
\end{eqnarray}
%
where $\Xi(h)=h(h-1)$.

Let us now outline the argument already used in  Refs.~\onlinecite{Dusuel04_3} and \onlinecite{Dusuel04_4} to compute finite-size scaling 
exponents. The $1/N$ expansion of any  physical quantity $\Phi$ considered here has a simple structure.  It consists of two contributions which are, respectively, regular (reg) and singular (sing) when $h$ approaches the critical point.  
Schematically, one has
%
\begin{equation}
  \Phi_N(h)=
  \Phi_N^\mathrm{reg}(h)+\Phi_N^\mathrm{sing}(h).
\end{equation}
%
Nevertheless, it is clear that no divergence can occur at finite $N$ for these quantites or its derivatives with respect to $h$, even at the critical point. This basic fact straightforwardly leads to the scaling exponents. Indeed, a close analysis of the singular part $\Phi_N^\mathrm{sing}(h)$ allows us, in the vicinity of $h_c$, to write it as follows:
%
\begin{equation}
  \Phi_N^\mathrm{sing}(h)\simeq 
  \frac{\Xi(h)^{\xi_\Phi}}{N^{n_\Phi}}
  \mathcal{F}_\Phi\left[N\Xi(h)^{3/2} \right],
\end{equation}
%
where $\mathcal{F}_\Phi$ is a function that only depends on the scaling variable $N\Xi(h)^{3/2}$. To compensate the singularity coming from $\Xi(h)^{\xi_\Phi}$, one thus must have 
$\mathcal{F}_\Phi (x) \sim x^{-2 \xi_\Phi /3}$ so that 
$\Phi_N^\mathrm{sing}(1) \sim N^{-(n_\Phi+2\xi_\Phi/3)}$. 
The scaling exponents corresponding to (\ref{eq:expansion_e0})-(\ref{eq:expansion_S1xS2x}) are listed  in Table 
\ref{tab:exponents}. 

We emphasize that here, we have approached the critical point from the symmetric phase ($h\rightarrow 1^+$). However, we could, as for the LMG model \cite{Dusuel04_4}, have reached it from the broken phase. 
As pointed out by Richardson \cite{Richardson77} who derived the exact solution in this regime, the $1/N$ developments of the ground-state energy and the occupation number (which is essentially the magnetization in the spin language)  are also singular at the critical point.  
We have checked that the scaling argument given above also predicts  
$e_0^\mathrm{sing}(N) \sim N^{-4/3}$ and $4\langle S_{1 z} \rangle^\mathrm{sing}/N \sim N^{-2/3}$ from this side of this transition, i.e., when $h\rightarrow 1^-$. 
Recently, the nontrivial scaling exponent of $e_0$ has been observed numerically \cite{Roman02} and analytically derived using a spin coherent state representation \cite{Roman04} but,  to our knowledge, the other scaling exponents given in Table  \ref{tab:exponents} have never been discussed.  
The present results can be compared to those recently obtained in the LMG model \cite{Dusuel04_3,Dusuel04_4}. For the isotropic LMG model, which has the same interaction term ($S_x^2+S_y^2$) as the Hamiltonian (\ref{eq:ham_bcs_spins}), scaling exponents are trivial since the $1/N$ expansion is not singular at the critical point. However, for the anisotropic LMG model, similar exponents are found (multiple of $1/3$) except  that $x$ and $y$ directions have different exponent.

\begin{figure}[t]
  \centering
  \includegraphics[width=8cm]{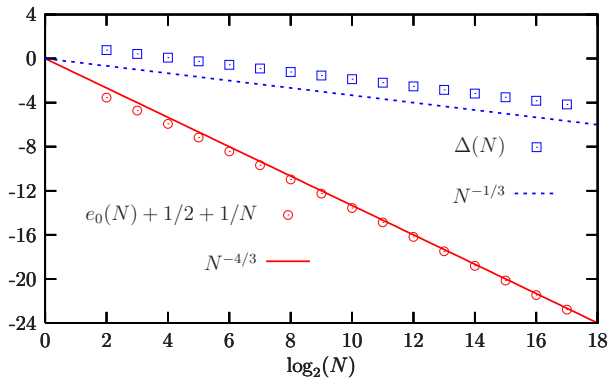}
  \caption{Behavior of the ground-state energy per spin $e_0$ and the gap $\Delta$ as a function of the system size $N$ 
  ($\log_2-\log_2$ plot) at the critical point compared to analytical results.}
  \label{fig:gse_and_gap}
\end{figure}
\begin{figure}[t]
  \centering
  \includegraphics[width=8cm]{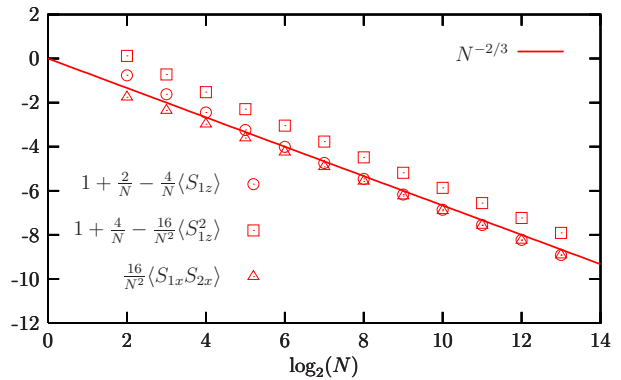}
  \caption{Finite-size scaling of the magnetization and the correlation functions ($\log_2-\log_2$ plot) at the critical point. A clear power-law behavior is observed with a nontrivial scaling exponent $2/3$ predicted by the present method.}
  \label{fig:observables}
\end{figure}

To check the validity of the present approach, we have performed numerical diagonalizations of critical finite-size systems, with up to $N=2^{17}$ spins for $e_0$ and $\Delta$ and up to $N=2^{13}$ spins for the magnetization and the correlation functions, which require the knowledge of the eigenstates. The results are depicted in Figs. \ref{fig:gse_and_gap} and \ref{fig:observables} and show an excellent agreement with the analytical predictions. Note that the regular part   $\Phi_N^\mathrm{reg}(1)$ has been substracted to underline the nontrivial scaling behavior.

%


Let us now discuss the entanglement properties of  the ground state.
Here we focus on the concurrence \cite{Wootters98}, which
characterizes the entanglement between two spins. As in the LMG
model \cite{Vidal04_1}, in the thermodynamical limit the ground
state becomes a completely separable state. Thus the nontrivial
properties of the concurrence are encoded in the finite $N$
corrections and one has to consider the rescaled concurrence
$C_\rmR=(N-1) C$. For the two-level reduced BCS model, one has to
distinguish between two cases: {(\it i)} both spins belong to the
same subset ($I_1$ or $I_2$); {(\it ii)} each spin belongs to distinct
subsets. In both situations, one can express the rescaled
concurrence as a function of the observables previously calculated.
Generalizing for our purpose the results given in Ref.~\onlinecite{Wang02}, 
one gets
%
\begin{eqnarray}
  \label{eq:conc_def}
  C_\rmR^{1,1}&=&C_\rmR^{2,2}=2\max(0,w-\sqrt{v_+ v_-}),\\
  C_\rmR^{1,2}&=&C_\rmR^{2,1}=2(y-v).
\end{eqnarray}
%
The superscript refers to the subset to which the spins considered belong to, and 
%
\begin{eqnarray}
  w &=& \frac{1}{4}\left[ 1-\frac{N}{N-2}
    \left( \frac{16\langle S_{1z}^2 \rangle}{N^2}
      -\frac{2}{N}\right)\right],\\
  v_\pm &=& \frac{1}{4}\left[ 1 \pm 2\frac{4\langle S_{1z} \rangle}{N}
    + \frac{N}{N-2} \left( \frac{16\langle S_{1z}^2 \rangle}{N^2}
      -\frac{2}{N}\right)\right],\quad\;\;\\
  y &=& \frac{1}{2}\frac{16\langle S_{1x}S_{2x} \rangle}{N^2}, \quad\quad 
  v = \frac{1}{4}\left( 1-\frac{16\langle S_{1z}^2 \rangle}{N^2} \right).
\end{eqnarray}
%

\begin{figure}[t]
  \centering
  \includegraphics[width=8cm]{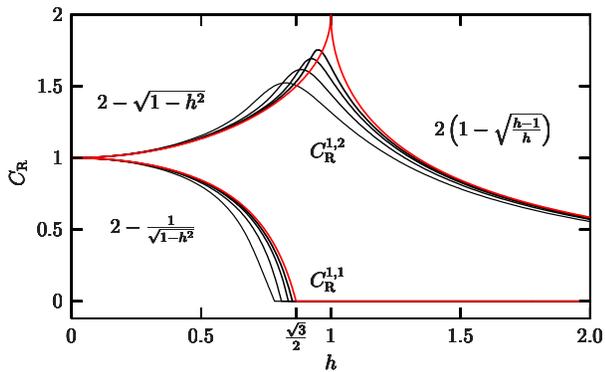}
  \caption{Rescaled concurrence of the ground state as a function
  of the magnetic field for $N=32$, 64, 128, 256, and $\infty$.
  In the thermodynamical limit a cusplike singularity appears at the critical point.}
  \label{fig:conc_vary_N}
\end{figure}

Using expressions (\ref{eq:expansion_S1z})-(\ref{eq:expansion_S1xS2x})  truncated to order $(1/N)^1$,  one can compute the thermodynamical limit of the rescaled concurrence in the  symmetric phase $(h \geq 1)$, and one finds 
%
\begin{equation}
  \label{eq:conc_sym}
  C_\rmR^{1,1} = 0, \quad \quad C_\rmR^{1,2} = 2 \left(1-\sqrt{h-1 \over h} \right).
\end{equation}
%
In the broken phase $(0\leq h \leq 1)$, one has to use the 
Holstein-Primakoff representation around the mean-field ground state. 
We found out that it is then enough to perform a Bogoliubov transform 
to compute the concurrence. Indeed, although such a simple calculation 
fails to provide the full $(1/N)^1$ contribution to the spin expectation 
values (as explained in Ref.~\onlinecite{Dusuel04_4} for the LMG model), 
the unknown parts of the $(1/N)^1$ contributions cancel each other when 
computing the thermodynamical limit of the rescaled concurrence, and one 
obtains
%
\begin{eqnarray}
  \label{eq:conc_brok}
  C_\rmR^{1,1} &=& \max\left(0\,,\,2 - {1 \over \sqrt{1- h^2 }}\right),\nonumber\\
  C_\rmR^{1,2} &=& 2 -  \sqrt{1- h^2 }.
\end{eqnarray}
%
The rescaled concurrence is depicted  in Fig. \ref{fig:conc_vary_N} for several system sizes as well as in the  thermodynamical limit. Let us also mention that the finite-size scaling exponent for $C_\rmR^{1,2}$ can be computed using the same scaling argument as previously and equals $1/3$ instead of $2/3$ as expected from the scaling of the observables.

As in the LMG  or the Dicke model \cite{Lambert04}, it is
interesting to note that $C_\rmR^{1,2}$ displays a cusplike
behavior at the critical point, whereas it is a smooth function for
$h \neq 1$. More interestingly, the max function used in definition
(\ref{eq:conc_def}) leads to $C_\rmR^{1,1}=0$ for $h\geq\sqrt{3}/2$.
It is worth noting that this value of the magnetic field does not
play any special role in the phase diagram, whereas it naturally
arises in the entanglement analysis. However, if we do not consider
the max function, $C_\rmR^{1,1}$ diverges at the critical
point.
In the zero-field limit, the ground state is simply the Dicke state
corresponding to $S=N/2$ and $S_z=0$, whose rescaled concurrence
equals 1 \cite{Wang02,Stockton03}. Note that, in this limit, the
distinction between subsets $I_1$ and $I_2$ becomes irrelevant so that both
rescaled concurrences are identical. In the infinite $h$ limit, the
ground state is the separable state 
$|\psi_0\rangle=(\otimes_{i \in I_1} |\!\!\uparrow \rangle_i)
\otimes(\otimes_{j \in I_2} |\!\!\downarrow \rangle_j)$, 
which has also a vanishing total magnetization ($S_z=0$) 
but which is not an eigenstate of ${\bf S}^2$. 
However, it is clear that the concurrence of such a state is
exactly zero whatever the two considered spins. 


The method used in the present work is certainly well suited to tackle many similar problems for which a semiclassical description of the thermodynamical limit is exact as illustrated here or in the LMG model. An interesting issue is to understand the key ingredients that allow for exact solutions of the flow equations. In particular, it would be of special interest to analyze, along the same line,  models where some regions of the phase diagram are known to be chaotic and others integrable. 
Finally, let us note that our method may also be useful to compute the von Neumann entropy which displays some interesting features in quantum critical systems \cite{Latorre03,Latorre04_1,Latorre04_3,Its04}.\\

\acknowledgments

We are indebted to J. Dukelsky, A. Reischl, A. Rosch and K.~P. Schmidt for fruitful and valuable discussions. We also warmly thank G. Sierra for sending us unpublished results about the critical properties of the BCS model \cite{Roman04}. S. Dusuel  gratefully acknowledges financial support of the DFG in SP1073.


\end{document}